# The Quest for the Ideal Scintillator for Hybrid Phototubes

Bayarto K. Lubsandorzhiev, Bruno Combettes

*Abstract*—In this paper we present the results of extensive studies of scintillators for hybrid phototubes with luminescent screens. The results of the developments of such phototubes with a variety of scintillators are presented. New scintillator materials for such kind of application are discussed. The requirements for scintillators to use in such hybrid phototubes are formulated. It is shown that very fast and highly efficient inorganic scintillators like ZnO:Ga will be ideal scintillators for such kind of application.

## I. INTRODUCTION

EXPERIMENTAL neutrino astrophysics has been developing with impressive pace for the last three decades. The detection of neutrinos from the SN1987A supernova burst and the discovery of neutrino oscillation are tremendous scientific achievements. One can assign this success to an extent to the development of big hemispherical vacuum photodetectors widely used in the overwhelming majority of experiments in the field. The most popular photodetectors are classical photomultiplier tubes (PMTs). Unfortunately PMTs have a number of shortcomings: poor collection and effective quantum efficiencies, poor time resolution, prepulses, late pulses, afterpulses, susceptibility to terrestrial magnetic field. In many respects vacuum hybrid phototubes are free from above mentioned disadvantages. Vacuum hybrid photodetectors are Hybrid Photodiodes (HPDs), Hybrid Avalanche Photodiodes (HAPDs) and Hybrid Phototubes with luminescent screens. HPD and HAPD are vacuum photodetectors using silicone diodes and avalanche diodes correspondingly for photoelectron multiplication. They have excellent characteristics and at least one but substantial drawback – they are very expensive because a transfer technique is used in the manufacturing process and last but not least their gain is still rather low.

In the early 1980s G. van Aller, S.-O. Flyckt and W.Kuhl from PHILIPS Laboratory at that time (now PHOTONIS Group) had developed the "smart" 15" phototube XP2600 [1] which was the first hybrid phototube with a luminescent screen. The tube had been developed for the first deep underwater muon and neutrino detector (DUMAND) project [2] and successfully tested in Lake Baikal [3]. Especially for another pioneering experiment – the lake Baikal neutrino experiment the 16" QUASAR-370 phototube [4-7] has been developed in Russia by a close collaboration of Institute for Nuclear Research in Moscow and KATOD Company in Novosibirsk, following the basic design of the "smart" phototube. The Baikal neutrino detector is the first full scale deep underwater neutrino telescope in the world [8]. It is being still operated with more than 200 QUASAR-370 phototubes in the array [9].

Both phototubes suit very well to the requirements for photodetectors of large-scale Cherenkov detectors. They are fast, highly sensitive to Cherenkov light and immune to terrestrial magnetic field. They have no prepulses and late pulses. Afterpulses of the phototubes are strongly suppressed in comparison to conventional PMTs [10]. It will be shown in the next chapters of the paper that the parameters of scintillators used in the luminescent screens of the phototubes are of utmost importance for the parameters of the phototubes.

## II. HYBRID PHOTOTUBE WITH LUMINESCENT SCREEN

### A. Operational principle

The hybrid phototube with a luminescent screen is a combination of an electro-optical preamplifier with a large hemispherical photocathode and small conventional PMT. The QUASAR-370 phototube's drawing is presented in figure 1. Photoelectrons produced by incident photons on the large hemispherical photocathode of the optical preamplifier with > $2\pi$ viewing angle are accelerated by high electric field of ~25 kV and hit a luminescent screen which is fixed near the center of the glass bulb. The luminescent screen is a thin layer of a fast, high light yield scintillator covered by an aluminium foil. Light flashes produced by photoelectrons in the scintillator are registered by a small PMT. As a result one photoelectron from the optical preamplifier engenders 20-30 photoelectrons in the small PMT. The combination of the luminescent screen and the small PMT's photocathode can be likened to the first stage of conventional PMTs dynode system. This high gain first

B. K. Lubsandorzhiev is with the Institute for Nuclear Research of Russian Academy of Sciences, 117312, Moscow, Russia (corresponding author to provide phone: +791611483810; fax: +74951352268; e-mail: lubsand@pcbai10.inr.ruhep.ru).

B. Combettes is with Photonis Group, 19106 Brive Cedex, France, (e-mail: b.combettes@photonis.com).

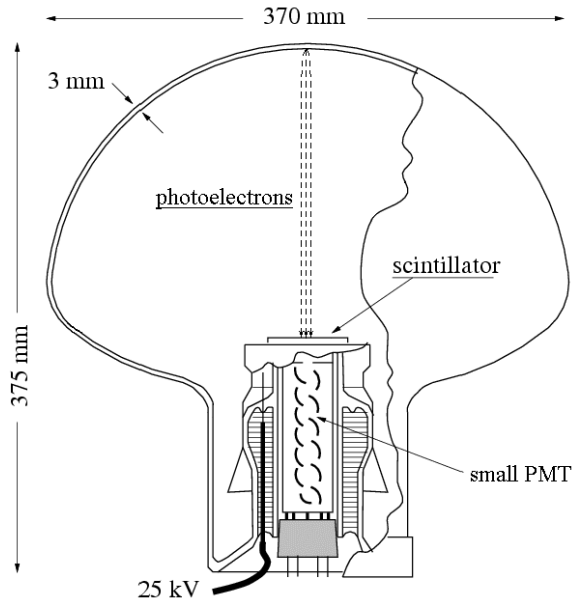

Fig. 1. The Quasar-370 hybrid phototube

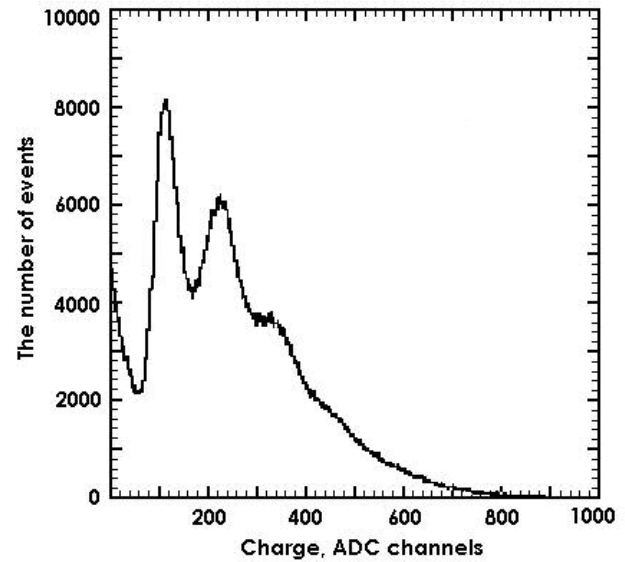

Fig. 2. Charge distribution of multi photoelectron pulses of the QUASAR-370 phototube with YSO scintillator.

stage results in an excellent single photoelectron resolution and together with high accelerating voltage allows to keep low the time jitter of the phototube. The QUASAR-370 phototube has a nonspherical mushroom shape of its glass bulb to provide more isochronic photoelectron trajectories. The maximum transit time difference of photoelectrons is less than 1 ns.

*B. Single photoelectron resolution of QUASAR-370.*

The single photoelectron resolution of the QUASAR-370 phototube is defined mainly by the gain $G$ of the electro-optical preamplifier. $G$ is the ratio of the number of photoelectrons $N_{p.e.\_pmt}$ detected by the small PMT to the number of photoelectrons $N_{p.e.\_preamp}$ emitted at the preamplifier photocathode:

$$G = N_{p.e.\_pmt} / N_{p.e.\_preamp} \quad (1)$$

$Y(E_e)$ – scintillator's light yield as a function of photoelectron kinetic energy $E_e$, $\xi$ - light collection efficiency, $\eta$ - effective collection efficiency of the small PMT.

Fig.2 shows typical charge distribution for a few photoelectrons pulses of QUASAR-370. The high amplification factor of the first stage allows to separate pulses of one, two, three and identify even a hump due to four photoelectron events. One can conclude that the higher level of the first stage gain $G$ the better single photoelectron resolution.

*C. Timing of QUASAR-370.*

A single photoelectron pulse of QUASAR-370 is a superposition of $G$ single photoelectron pulses of the small PMT, distributed exponentially in time:

$$P(t) = (1/\tau)exp(-1/\tau) \quad (2)$$

with $\tau$ being the decay time constant of the scintillator.

The best time resolution (jitter) is reached by using a double threshold, "slow-fast" discriminator [4]. It consists of two discriminators with different thresholds and integration constants: a timing, "fast", discriminator with a threshold of $0.25q_1$ and a strobing, "slow", discriminator with a threshold of $0.25Q_1$ ($q_1$ and $Q_1$ are the mean charges of single photoelectron pulses of the small PMT and the electro-optical preamplifier respectively).

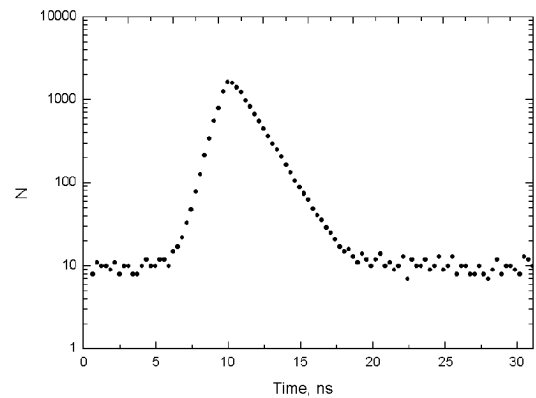

Fig. 3. Single photoelectron transit time distribution of the QUASAR-370 phototube.

Events arrival time is defined by the front edge of the first of the $G$ single photoelectron pulses of the small PMT. Typical single photoelectron transit time distribution of QUASAR-370 measured by such technique is shown in fig. 3. In this case, the transit time distribution of single photoelectron pulses of the whole QUASAR-370 phototube is described well by the following expression:

$$W(t) \sim exp(-(G/\tau)t) \quad (3)$$

$W(t)$ is determined by the ratio between the scintillator decay time constant $\tau$ and the gain $G$. Strictly speaking (3) describes precisely just the right but the main part of the distribution. The left part is defined mainly by a scintillation rise time and jitters of the small PMT and electronics. As it is clearly seen in fig. 3, QUASAR-370 has no prepulses and late pulses. The reason for the lack of prepulses is the fact that the first stage of the phototube is optically isolated from the phototube's cathode. Moreover the complete vacuum separation between electro-optical preamplifier and the small PMT leads to the much lower level of afterpulses in QUASAR-370 in comparison with conventional PMTs. Late pulses are suppressed due to the high gain of the first stage. One can find more details of the phototube timing and single photoelectron response and photoelectron backscattering in [4-7, 15]. So, the higher the gain $G$ and shorter decay time $\tau$ of the scintillator the better time resolution of the phototube.

*D. Luminescent screen*

The scintillator for the basic design of the QUASAR-370 phototube was chosen to be YSO ($Y_2SiO_5$:Ce). This scintillator was selected because it was the most effective, fast and inorganic scintillator among available scintillators at that time. It was used as a phosphor or a monocrystal in the phototube. The phosphor thickness is 6 μm which is optimized for the effective detection of 25 keV photoelectrons. It is deposited on 20 mm in diameter glass wafer. The monocrystal is ~100 μm thick just for mechanical stability and 20 mm in diameter. An aluminium foil of 100 nm thickness covers the phosphor or the monocrystal providing complete optical separation of the scintillator and the phototube's photocathode. The aluminium foil serves as a reflector too increasing effectively the light yield of the luminescent screen of the phototube. The latter is about 25% NaI(Tl) in average and its decay time is in the range of 30-50 ns. In the next chapter the results of extensive developments of new luminescent screens with new scintillators will be described.

## III. SCINTILLATORS FOR HYBRID PHOTOTUBES

The first stage gain $G$ of the phototube is directly connected with the scintillator light yield. The equation (1) for the first stage gain can be rewritten in another way:

$$G = Y(E_e)\xi\eta_{eff} \quad (4)$$

$Y(E_e)$ - scintillator light yield, $E_e$ - photoelectron energy, $\xi$ - collection efficiency of light on the small PMT photocathode, $\eta_{eff}$ – effective collection efficiency of the small PMT.

Taking into account equations (1)-(4) we come to the conclusion that to have hybrid phototubes with better time and amplitude resolutions and faster time response one should search for scintillators with light yield as high as possible and at the same time decay time as short as possible. Unfortunately the nature does not provide us with wide opportunities. The reason is that scintillators should meet rather strong requirements:
1) Vacuum compatibility. The phototube is a vacuum device so the scintillator should be vacuum compatible.
2) Hardness to the phototube manufacturing process. It's very desirable to avoid transfer technique in the phototube manufacturing to keep prices low. So the scintillator should withstand photocathode manufacturing processes: high temperature, aggressive chemical atmosphere etc.
3) High light yield for electrons with energies of 10-30 kV.
4) Fast emission kinetics.
5) Low $Z_{eff}$ to suppress photoelectron backscattering effect.
6) Emission spectrum should suit the small PMT's photocathode sensitivity.
7) Good proportionality.

The first two requirements restrict scintillators to just nonhygroscopic, inorganic scintillators. In Table 1 some appropriate scintillators with their basic parameters are presented. More details on the most part of scintillators listed in the Table 1 can be found in [11-13] and in all references therein.

TABLE I
SCINTILLATORS APPROPRIATE FOR USE HYBRID PHOTOTUBES

| Scintillator | $Z_{eff}$ | $\lambda_{max}$ nm | Light Yield %NaI(Tl) | Decay time Ns |
|---|---|---|---|---|
| YSO ($Y_2SiO_5$:Ce) | 33 | 420 | 25 | 30-50 |
| GSO ($Gd_2SiO_5$:Ce) | 59 | 440 | 25 | 50-60 |
| YAP ($YAlO_3$:Ce) | 36 | 390 | 40 | 25-30 |
| LuAP ($LuAlO_3$:Ce) | 65 | 360 | 30-40 | 20 |
| LPS ($Lu_2Si_2O_7$:Ce) | 64 | 380 | 70-80 | 30 |
| LSO ($Lu_2SiO_5$:Ce) | 66 | 420 | 75 | 40 |
| SBO ($ScBO_3$:Ce) | 17 | 380 | 40 | 30 |
| $BaF_2$ | 54 | 220 320 | 6 25 | 0.8 630 |
| LS ($Lu_2S_3$:Ce) | 67 | 590 | 80 | 30 |
| ZnO:Ga | 28 | 440 | 40-100 | 0.4 |

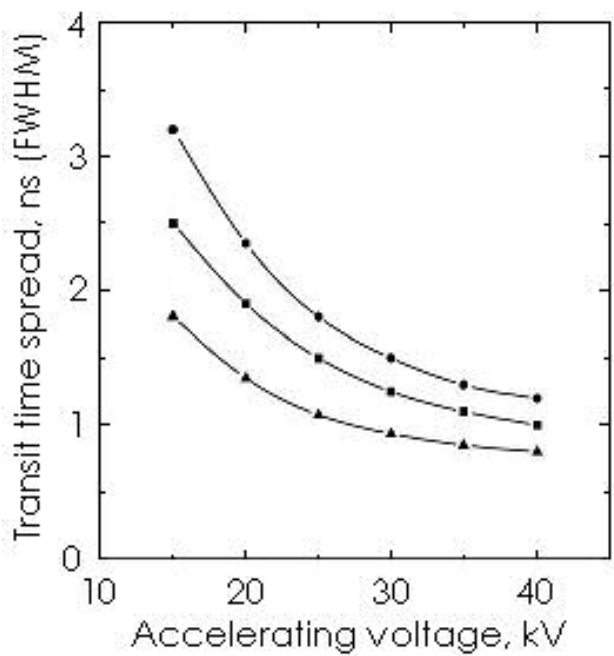

Fig. 4. Dependence of photoelectron transit time spread (FWHM) of the QUASAR-370 phototube with different scintillators on accelerating voltage. ● - YSO, ■ - SBO and YAP, ▲ - LSO.

For the last 15 years a number of modifications of the QUASAR-370 phototube have been developed at INR in close cooperation with KATOD Company. For more details we refer to [14-16]. The QUASAR-370 phototube's modifications were equipped with luminescent screens with different scintillators: SBO ($ScBO_3$) [17], YAP, LSO among them. The scintillators were in the form of phosphors and monocrystals. The dependencies of photoelectron transit time spread (FWHM) for the QUASAR-370 phototubes with different scintillators on accelerating voltage are shown in fig. 4.

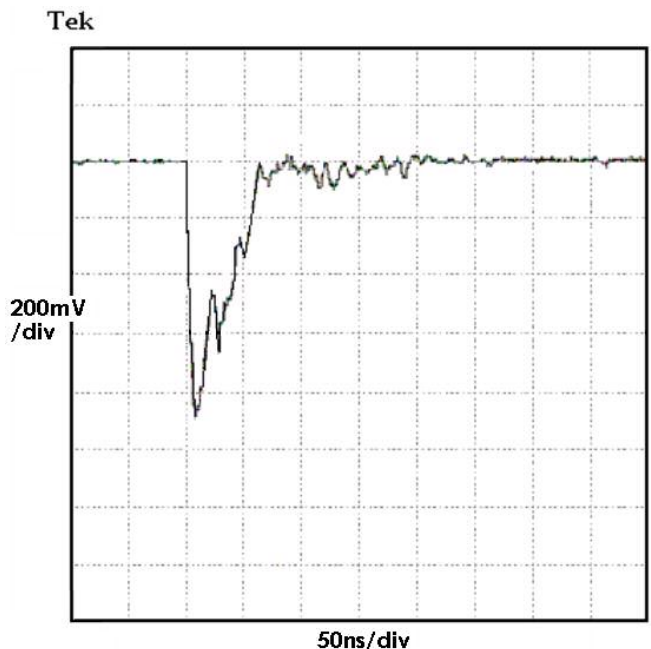

Fig. 5. Output signal waveform of the QUASAR-370 phototube.

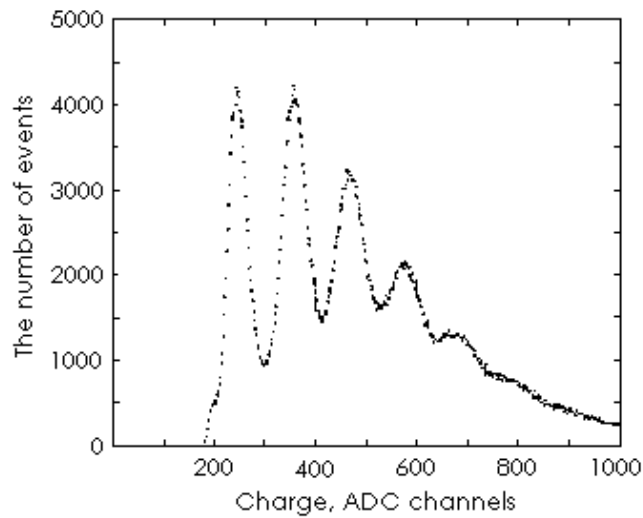

Fig. 6. Charge distribution of multi photoelectron pulses of the QUASAR-370 phototube with LSO monocrystal scintillator.

So far the best results has been reached with LSO monocrystal scintillator despite its nonproportionality problems and high $Z_{eff}$. The latter doesn't play crucial role in the phototube's time resolution, albeit it is important in the phototube's time response. A typical wave form of the QUASAR-370 phototube's output signal is depicted in fig.5. The phototube was illuminated by short ~0.5 ns width (FWHM) light pulses from $N_2$ laser. One can see rather sharp peaks due to photoelectrons backscattered on the luminescent screen. As for the photoelectron transit time spread of the phototube it is about 1 ns (FWHM) for 25 kV accelerating voltage, see fig. 4. Presently 25 kV can be considered as a safety region of a stable operation of the phototube. Single photoelectron resolution of the phototube with LSO scintillator is about 35% (FWHM). Fig. 6 presents the charge distribution of multi photoelectron pulses of the QUASAR-370 phototube with LSO monocrystal scintillator. Distinct peaks due to up to 7 photoelectrons are clearly seen. A cut-off in the left side of the spectrum is due to a discriminator threshold.

IV. FIGURE OF MERIT

It is very convenient for such kind of application where scintillator's timing and light yield of utmost importance to define a figure of merit $F$ of scintillators:

$$F = (Y/\tau)ab \qquad (5)$$

$Y$ – scintillator's light yield; $\tau$ - scintillator's decay time; $a$ – a coefficient accounting for a scintillation light detectability by the small PMT; $b$ – a coefficient connected with compatibility of scintillator with vacuum and manufacturing process of the phototube. In fact, the coefficient $a$ describes how well the scintillator's emission spectrum matches to the small PMT sensitivity and transparency of the phototube's glass bulb. The coefficient $b$ equals to zero for organic and plastic and hygroscopic scintillators. For other scintillators: $b = 1$.

In Table 2 the figures of merit for the best scintillators for hybrid phototubes are presented. Here LS is $Lu_2S_3$:Ce scintillator [18], so called "red" scintillator, with $\lambda_{max}$ = 590 nm. The scintillator is interesting in case of readout by a silicone photomultiplier (SiPM) or a photodetector with A3B5 photocathode with higher sensitivity at longer wave lengths in comparison with conventional PMTs.

TABLE I
FIGURE OF MERIT FOR SCINTILLATORS

| Scintillator | YSO | YAP | SBO | LSO | LS | ZnO |
|---|---|---|---|---|---|---|
| $F$ | 1 | 1.3 | 1.3 | 1.8 | 4[*] | 250 |

[*] - the value of $F$ in this case is evaluated taking into account SiPM readout instead of PMT readout.

The best scintillator is ZnO:Ga with light yield of 40÷100%NaI(Tl) and 0.4 ns decay time. There is a variety of data on this scintillator in literature [19-21]. Its emission spectrum has maximum at 440 nm, it allows to use ordinary, cheap PMTs. The figure of merit for ZnO:Ga is a huge, exceeding other scintillators by nearly two orders of magnitude. Unfortunately so far all attempts to produce the QUASAR-370 phototube with luminescent screen based on ZnO:Ga scintillator have not been successful. But still this scintillator would be almost ideal scintillator for hybrid phototubes with luminescent screens and it is of foremost importance to continue studies of the scintillator.

## V. CONCLUSION

Vacuum hybrid phototubes with luminescent screens have excellent time and amplitude resolutions even with existing scintillator (YSO). Modifications of QUASAR-370 phototube with faster and more efficient scintillators (e.g. LSO, SBO and YAP) demonstrated quite clearly that such phototubes keep well ahead of other vacuum photodetectors including HPDs. Very fast and efficient ZnO:Ga scintillator is almost the ideal scintillator for such kind of application. If we will manage to have good reliable ZnO:Ga scintillator, we will have the ideal photodetector for the next generation of giant neutrino and other challenging experiments of 21$^{st}$ century.

ACKNOWLEDGMENT

The authors dedicate the paper to the memory of their late friend S.-O. Flyckt. He has been widely renowned in physics community. The whole community called him just Esso. As a co-inventor of the "smart" phototube Esso was great champion of hybrid phototubes with luminescent screens and ardent supporter of all activities concerning scintillator developments for such phototubes. Esso would be happy to know that Photonis Group is going to revive the "smart" phototube development at new level.

The authors are indebted to Dr. V.Ch. Lubsandorzhieva for careful reading of the manuscript and many invaluable remarks.

REFERENCES

[1] G. van Aller, S.-O. Flyckt, W. Kuhl. "An electro-optical preamplifier combination with integrated power supply offering excellent single electron resolution for DUMAND," *IEEE Trans. Nucl. Sci.*, vol. 30, pp. 469-474, 1983.
[2] W. Andreson, T. Aoki, H.G. Berns et al. "DUMAND II – Proposal to Construct a Deep-Ocean Laboratory for the study of High-Energy Neutrino Astrophysics and Particle Physics," *HDC-2-88, University of Hawaii*, 1988.
[3] L.B. Bezrukov, B.A. Borisovets, A.V.Golikov et al. "Properties and test results of a photon detector based on the combination of electro-optical preamplifier and a small photomultiplier," in *Proc. 2$^{nd}$ Intern. Symp. on Underground Physics,* Baksan Valley, pp. 230-236, 1987.
[4] R.I. Bagduev, L.B. Bezrukov, B.A. Borisovets et al. "The Optical Sensor of the Lake Baikal Project," in *Proc. 2$^{nd}$ Intern. Conf. on Trends in Astroparticle Physics.* Aachen, pp. 132-137, 1991.
[5] R.I. Bagduev, L.B. Bezrukov, B.A. Borisovets et al. "Highly sensitive fast photodetector Quasar-370 for large-scale experiments in cosmic ray physics," *Bull. Of the Russian Academy of Sciences: Physics,* vol. 57, no. 4, pp. 135-139, 1993.
[6] L.B. Bezrukov, B.A. Borisovets, A.A. Doroshenko et al. "QUASAR-370 – The Optical Sensor of the Lake Baikal Neutrino Telescope," in *Proc. 3$^{rd}$ NESTOR Intern. Workshop,* Pylos, pp. 645-658, 1993.
[7] R.I. Bagduev, L.B. Bezrukov, B.A. Borisovets et al. "The Optical Module of the Baikal Deep Underwater Neutrino Telescope," *Nucl. Instrum. Meth. A,* vol. 420, pp. 138-152, 1999.
[8] I.A. Belolaptikov, L.B. Bezrukov, B.A. Borisovets et al. "The Baikal underwater Neutrino Telescope," *Astroparticle Physics,* vol. 7, pp. 263-278, 1997.
[9] V. Aynutdinov, V. Balkanov, I. Belolaptikov et al. "The Baikal Neutrino Experiment: From NT200 to NT200+," *Nucl. Instrum. Meth. A,* vol. 567, pp.433-437, 2006.
[10] B.K. Lubsandorzhiev, R.V. Vasiliev, Y.E. Vyatchin, B.A.J.Shaibonov, "Photoelectron backscattering in vacuum phototubes," *Nucl. Instrum. Meth. A,* vol. 567, pp. 12-16, 2006.
[11] S.E. Derenzo, M.J. Weber, E. Bourret-Courchesne, M.K. Klintenberg. "The quest for the ideal inorganic scintillator," *Nucl. Instrum. Meth. A,* vol. 505, pp.111-117, 2003.
[12] C.L. Melcher. "Perspectives on the future development of new scintillators," *Nucl. Instrum. Meth. A,* vol. 537, pp. 6-14, 2005.
[13] P. Dorenbos. "Light output and energy resolution of $Ce^{3+}$–doped scintillators," *Nucl. Instrum. Meth. A,* vol. 486, pp. 208-213, 2002.
[14] B.K. Lubsandorzhiev, "Development of hig sensitive Light Detectors for Underwater Neutrino Telescopes," in *Proc. 25$^{th}$ ICRC,* Durban, vol. 4, pp. 129-132., 1997.
[15] B.K. Lubsandorzhiev, "Photodetectors of the lake Baikal Neutrino experiment and TUNKA Air cherenkov Array," *Nucl. Instrum. Meth. A,* vol. 442, pp. 368-373, 2000.
[16] B.K. Lubsandorzhiev, "New developments of photodetectors for the lake Baikal neutrino experiment," in *Proc. ICATPP-7,* Como, pp. 79-84, 2001.
[17] N.P. Soshin. "Adjsutment of emission characteristics and energy yield of industrial phosphors," *Bull. of USSR Academy of Sciences: Physics*, vol. 43, pp. 6-9, 1979.
[18] J.C. van't Spijker, P.Dorenbos, C.P. Allier et al. "$Lu_2S_3$:$Ce^{3+}$, A new red luminescing scintillator," *Nucl. Instrum. Meth. B,* vol. 134, pp.304-309, 1998.
[19] D. Luckey. "A fast inorganic scintillator," *Nucl. Instrum. Meth.,* vol. 62, no. 1, pp.119-123, 1968.
[20] S.E. Derenzo, M.J.Weber, M.K.Klintenberg. "Temperature dependence of the fast, near-band-edge scintillation from CuI, $HgI_2$, $PbI_2$, ZnO:Ga and CdS:In," *Nucl. Instrum. Meth. A,* vol. 486, pp. 214-219, 2002.
[21] P.J. Simpson, R. Tjossen, A.W. Hunt, K.G. Lynn, V. Munne. "Superfast timing performance from ZnO scintillators," *Nucl. Instrum. Meth. A,* vol. 505, pp. 82-84, 2003.